\documentclass[twocolumn,10pt]{IEEEtran}
\usepackage{cite}
\usepackage{graphicx}
\usepackage{url}
\usepackage{amsmath}
\usepackage{latexsym,epsfig,amssymb}

\newcommand{\BEQA}{\begin{eqnarray}}
\newcommand{\EEQA}{\end{eqnarray}}

\newtheorem{theorem}{Theorem}

\newtheorem{corollary}[theorem]{Corollary}

\begin{document}
%\def\baselinestretch{0.9}

\title{Peer to Peer Networks for \\Defense Against Internet Worms}

\author{Srinivas Shakkottai , \thanks {Dept. of Electrical and Computer
Engineering, and
Coordinated Science Laboratory, University of Illinois at
Urbana-Champaign,
%Urbana, Illinois 61801\\
Emails: sshakkot,rsrikant@uiuc.edu}
\and R.~Srikant 
%\date{\small{Emails: {\tt
 %     (sshakkot,rsrikant)@uiuc.edu}}}
}
\maketitle

\begin{abstract}
Internet worms, which spread in computer networks without human mediation,
pose a severe threat to computer systems today.  The rate of propagation
of worms has been measured to be extremely high and they can infect a
large fraction of their potential hosts in a short time.  We study  two
different methods of patch dissemination to combat the spread of worms.  We
first show that using a fixed number of
patch servers performs woefully inadequately against Internet worms.  We
then show that by exploiting the exponential data dissemination capability
of P2P systems, the spread of worms can be halted very
effectively. We compare the two methods by using fluid models to compute two
quantities of interest: the time taken to effectively combat the progress of
the worm and the maximum number of infected hosts.  We validate our models using Internet measurements and simulations.
\end{abstract}

%\category{H.1.0}{Information Systems}{Models and Principles}
%A category including the fourth, optional field follows...
%\category{D.2.8}{Software Engineering}{Metrics}[complexity measures, performance measures]

%\terms{Performance}
%\vspace{.2in}

%\noindent \textbf{Keywords} \\
%Worm Propagation, Peer-to-Peer Networks, Fluid Models, Patch Dissemination, Monitoring

\section{Introduction}

The advent of malicious mobile code has lead to a paradigm shift in Internet security applications.  Earlier, computer viruses were inherently limited by the fact that human mediation was required for them to propagate, which also meant that human intervention was sufficient to contain them.  However, with increased connectivity of computers and availability of information regarding vulnerabilities of operating systems and applications, there have been several instances of malicious code that propagate on their own.   Such mobile malicious code are now called \emph{worms}. Interest in worms has been fueled by headline-making attacks causing near cessation of Internet services, and the names of these worms -- such as Code-Red, Slammer and Blaster-- are now known to most Internet users.

\begin{figure}[htbp]
\begin{center}
\epsfig{file=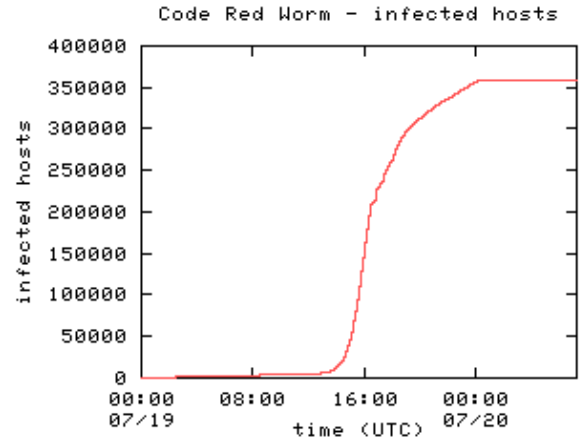,width=3.2in} \caption{Graph from \cite{mooshabro02} based on Internet measurement data, illustrating the nature of  propagation of the Code-Red (v2) worm.}
\label{fig_caida_worm}
\end{center}
\end{figure}

Measurement studies indicate that worm propagation usually follows the classical sigmoid curve as illustrated in Figure \ref{fig_caida_worm}.  The figure, which is obtained from  \cite{mooshabro02}, shows the propagation of the Code-Red (v2) worm measured over the duration of 24 hours.  There is an exponential growth stage followed by a slow finish stage.  The worm was programmed to switch from an `infection phase' to an `attack phase', and begin an attack on certain websites at a preselected time.  Such behavior is by no means unique to Code-Red.
The same kind of infect-then-attack behavior was observed with the Blaster worm \cite{baicoojah05} as well.  The Witty worm deleted small sections of the hard-disk contents on the infected hosts and its effects on the system were noticeable only over time \cite{weaell04}.  The Slammer worm was the most benign to the infected host -- all it did was to spread \cite{moopaxsav03}.  However, it caused storms of packets that overloaded networks as it spread.

Worms seen so far have not significantly injured their hosts during the time that they spread, since killing their host would prevent them from spreading effectively.  A host might actually be unaware that it is infected, as none of its functions are impaired.  This fact  means that one can deal with worm infestations by patching.  Hosts that are susceptible to the worm as well as those already infected could download and install a patch, which has the dual role of eliminating the malicious code from the host and closing the hole that enabled the infection in the first place. Typically patches are issued by either the creator of the OS or by a dedicated anti-virus provider. However, given the alarming rate at which worms can propagate (the Slammer worm infected more than 90 percent of vulnerable hosts within 10 minutes \cite{moopaxsav03}), there has been a need to rethink strategies for handling worm attacks. By and large, research has focused on three areas -- monitoring of worms, cutting down the rate of propagation (throttling) and delivering patches.

In our model we have a network of susceptible hosts that subscribe to the services of a patch provider. This assumption is made on the basis of the fact that major OS creators automatically provide subscription to their patching services. We assume that the number of infected hosts when the patch is released is small as compared to the total number of hosts, which is in accord with the fact that so far most attacks have happened after a vulnerably has been disclosed.  Also, worms which exploit previously unknown vulnerabilities (zero-day worms) have not been common \cite{lilnic04}. Once the patch is released, the provider sends an update message (which is tiny as compared to the patch) to all hosts which proceed to try and download the patch. 
%Both susceptible and infected hosts may be patched.  

When designing a system for the containment of worms, the main question which comes to mind is that of how long one has before the worm goes into `attack phase'.  If this time is sufficiently large, one could hope to patch a large fraction of computers before the attack occurs.  Considering the fact that worms have a stage in which their growth rate is exponential, even if the worm is slowed down, the time taken to infect a large fraction of hosts is likely to be small.   In such a case it is very possible that a fixed number of patch servers would be unable to cope with the spread of the worm.  It might then be advisable to combine throttling with a peer-to-peer (P2P) network that would be used for patch dissemination.

\subsection*{Related Work}

While the science of epidemiology or the study of causes, distribution, and control of disease in populations has been of interest to mankind for centuries, the past couple of years have seen an large upsurge of interest in the field. Interest in the area has stemmed from both computer worm epidemics as well as the organic kind.  There is now an ever increasing body of literature dealing with the measurement, modeling and analysis of computer worm propagation and prevention.  We highlight some important contributions in this area.

A good deal of work has gone into measuring the spread of worms on the Internet \cite{mooshabro02,moopaxsav03,baicoojah05}.  Researchers often try to reverse engineer the worm to understand its nature and the signature of its attack process.  The actual measurement is done by means of a \emph{network telescope}.  The idea here is to monitor a large fraction of the Internet address space \cite{zougontowgao05, mooshavoe04}.  Abnormal activity would register hits on the monitored space.  %Ideas on how large an address space ought to be monitored are present in .
In \cite{san03,coscro05}, there are ideas on how a P2P network could be used for monitoring of abnormal behavior.

Using simple fluid models \cite{kurtz78}, it is possible to study disease propagation using simple deterministic differential equations \cite{fra80,dalgan99}.   In the area of computer worms, initial work \cite{zougontow02, stapax02, chegaokwa03} largely focused on showing that the epidemiology model also applies to the spread of computer worms.  Basic study of defense systems is also present in this work.  More recently,  advanced models of worms, which include fine details such as non-uniform scanning rates, as well as ways to scale down the network for faster simulation that are accurate for certain worms like Slammer, have also been studied \cite{weaham04, keshamjiw05}.

Defense against worms, either by passive or active means, has developed in parallel with the worms themselves.
%, and there has been an effort to create better protection mechanisms for networks.
As models of worm proliferation have matured, using such models to make predictions on the performance of worm containment schemes has gained popularity.  Some interesting examples of such work are \cite{lilnic04,vojgan05a}. However, they concentrate on the number of infected hosts at infinite time, rather than at the time at which the attack phase of the worm begins.  However, an they do not consider the case when infected hosts can be patched. %In \cite{vojgan05a}, there is also the assumption that an infected host can never be patched.  This assumption does not seem to be true for any of the worms seen so far, since an unpatchable host is in effect killed (as the user would have to reformat or reinstall software), hindering the proliferation of the worm.  
As observed in \cite{hamhart05}, worms seen thus far have usually been fairly benign initially to the infected host so as to spread quickly, which means that the system operations are not significantly compromised.

Worm activity in an infected computer can be inferred by the fact that they tend try to set up new connections at a high rate.   This behavior immediately suggests  a way of slowing down the spread of worms.  By slowing down the rate at which new connections are established, worm applications can be retarded.   This is the principle behind virus throttling  \cite{will02,will03}.  Thus, throttling a virus buys time in which a patch may be disseminated in the network.

P2P networks have been showing ever increasing popularity as a means of data dissemination.  Internet users are now quite familiar with the concept and are well aware of software like KaZaa and BitTorrent which implement the idea. Since these systems usually have a large number of users, fluid models may be used in understanding their performance. Work on modeling and analysis of such systems is present in \cite{vecyan03,qiusri04,masvoj05}.

The P2P idea for worm containment has been considered in earlier work.  In \cite{lilnic04a}, the authors consider several types of worm defense mechanisms, including patching with a fixed number of patch servers and different types of ``patching worms'' that duplicate the worm's behavior to disseminate patches.  Using a graph-theoretic model, they show that the patching worms would perform better than a fixed number of patch servers.  However, the improvement in performance due to the patching worms is not quantified using the graph model.  They also consider the epidemic differential equation models to quantify the number of peak scans in the system.  In \cite{lilnic04}, the authors conduct an extensive numerical comparison between the performance of patching worms and content filtering and conclude that the two methods have comparable effects only when content filtering covers 89\% of the hosts.  Along with the monitoring aspect, \cite{coscro05} also considers the P2P idea for propagating alerts generated by the peers themselves about possible worm infestations and perform detailed simulations on the fraction of hosts that such a system could save.
\cite{vojgan05a} contains an extensive analytical study of worm propagation and a cooperative P2P system for patching is considered.  However, the P2P idea is not thoroughly investigated in this work.

\subsection*{How is our work different?}

The object of our study is to obtain a fundamental insight into the propagation of worms under active defense.  We use the fluid models describing worm scanning and containment schemes and solve them to obtain closed-form solutions. Once we have the solutions, our focus is on the \emph{orders of magnitude} of parameters (such as worm propagation time, maximum number of infected hosts, and patching time) in the system.  We express our results in terms of three quantities --
%Since we are interested in the time-frames at which events happen and not in arbitrary measures like minutes or hours, we measure time in terms of the average time taken for one infection to occur (for Code-Red this was about $33$ minutes and for Slammer this was about $36$ seconds).  This artifice largely eliminates the dependence on constants. We convert back to ``real'' time units as required in examples.
\begin{enumerate}
\item The total number of hosts in the system $N$ (a large number)
\item The \emph{virulence} of the worm denoted by $\beta$ (infections per unit time), which is the maximum rate at which the worm can spread.
\item The ratio of the maximum rate of patch propagation to worm's virulence denoted by $\gamma$ (dimensionless).
\end{enumerate}
We present our main insights below, starting with a fairly obvious one that serves as a benchmark, and proceeding to less intuitive ones:

\begin{itemize}
\item \textbf{A well designed worm will spread in $\Theta(\ln N)$ time to a significant fraction of the hosts.}\\
%If the total number of susceptible hosts is $N$, we have a significant infected population in $\Theta(\ln N)$ time.
While this result seems intuitively clear from the well-known exponential phase of worm spreading, it provides a useful benchmark to compare different patching schemes. Essentially, it says that any action that is taken to contain worms would have to be done within a logarithmic time frame, since a smart worm would switch to the attack phase at this time.

For example, for a worm like Code-Red with a susceptible population of about $360,000$ hosts, and $\beta=1.8$ infections per hour \cite{stapax02}, the value of  $\frac{1}{\beta}\ln N$ is about $7$ hours.  So if a patching scheme does not patch most of the hosts in $7$ hours, it is practically useless at dealing with it.

\item \textbf{With a fixed number of patch servers, both the maximum number of infected hosts and the time taken to disinfect the system are $\Theta(N)$.}\\
We show that in the case of a fixed number of patch servers, the time at which the infection starts to decay is $\Theta(\ln N)$ and that the number of infected hosts is $\Theta(N)$ at this time.  So  a fixed number of patch servers has practically no effect on the spread of the worm until most of the hosts are infected.  We also show that the time taken to wipe out the infection is $\Theta(N)$, so it takes a very long time for the system to be free of worms.  $\gamma$ plays almost no role in the results.

In the Code-Red like worm example, if we rely on a fixed number of patch servers,  even if $\gamma=300$, in roughly $7$ hours we have an infected population of $200,000$.  It takes about $25$ hours to rid the system of the worm.

\item \textbf{In P2P system, $\Theta(N^{\frac{1}{\gamma}})$ is the maximum number of infected hosts and $\Theta(\ln N)$ is the time taken to disinfect the system.}\\
We show that using P2P patch dissemination, the time at which the infection starts decreasing is $\frac{1}{\beta \gamma} \ln(\frac{N}{\gamma\overline{P}})$, the maximum number of infected hosts is $\Theta(N^{\frac{1}{\gamma}})$ (or $\Theta(N)$ if $\gamma\leq 1$), and the time taken for the system to be worm free is $\frac{1}{\beta \gamma}\left(1 +\frac{1}{\gamma}\right) \ln(N)$. Thus, the infection hits its peak and vanishes in $\Theta(\ln N)$ time.  The value of $\gamma$ can be increased by throttling the worm.   For $\gamma>1$, even small increases have a profound effect on P2P systems -- for instance, a $\gamma$ of $2$ shows performance of a greatly superior order than a $\gamma$ of $300$ in the fixed number of patch servers scheme.

%The P2P system creates new servers as time progresses, it too has the property of exponential spread like the worm.
For the Code-Red like worm example, with $\gamma$ being $2$, the maximum number of infected hosts is of the order $1000$ and the infection both hits its peak and is wiped out in about $5$ hours -- a paradigm shift from the fixed number of servers case!

%Although the order results provide a clear metric on system performance, it must be remembered that they do not give exact values for numericals.  For precise answers one must use the closed form solutions of the differential equations that we provide.

\item \textbf{The number of hosts to be monitored in order to get reliable measurements is $\Theta(\frac{N}{\ln N})$}\\
We show that if we are to obtain information about the worm's presence before it spreads to very many of the hosts (which takes $\Theta(\ln N)$ time), i.e., if we would like to know that the worm is in the system by $\ln \ln N $ time, then we have to monitor $\Theta(\frac{N}{\ln N})$ of the hosts in the system.  We also show that the same order or higher of hosts must be monitored in order to get a reliable estimate of the number of infected hosts at any time.

In the Code-Red like worm case, this means that if we want to know about the presence of the worm in the time $1/\beta\ \ln \ln (360,000)= 1.4$ hours, we would have to monitor about $28,000$ susceptible hosts.  For a worm that has the whole of the (IPv4) Internet as its prey, with $2^{32}$ addresses, this number is about $2^{27}$ a phenomenally large number!  To monitor such a large number of hosts one would need either the active participation of network operators or of the hosts themselves as a P2P system in identifying anomalous traffic.
%This makes a case for P2P monitoring as proposed in \cite{san03}.
\end{itemize}
\subsection*{Organization of the Paper}

We begin the paper in Section \ref{worm_model} by reviewing a differential equation model for the uniform scanning worm on the lines of the classical epidemic model.  The model has been solved earlier, and using the solution we show the exponential spreading of the worm.   The models and results in the rest of the paper are original and form the main contribution.  In Section \ref{patch_dissemination} we construct an analytical model of the patching process.  We create models for both the fixed number of servers and the P2P case and solve them.  From the solutions we make predictions on the performance of the systems in dealing with worms.  In Section \ref{experiments} we provide both measured data and simulations illustrating the characteristics of the patching process.   We then move on to the problem of monitoring the system for worms in Section \ref{monitors}. Finally, we conclude with pointers to extensions in Section \ref{conclusions}.

\section{Worm Propagation Model}\label{worm_model}

We first review the simple epidemic model to understand worm propagation. Let the number of hosts in the network be $N$. We assume that all hosts are identical in operation and that until a host has been patched, it is vulnerable to a worm. Let the number of susceptible hosts at time $t$ be denoted by $S(t)$.  Similarly, let the number of infected hosts at time $t$ be denoted by $I(t)$.  Then we have that at any time $t$,
\BEQA\label{eqn_tot_hosts}
S(t)+I(t)=N.
\EEQA
We assume that an infected host scans the address space of the network uniformly.  This assumption follows from the fact that under our model all hosts are identical, and so are equally vulnerable to the worm.  %Most worms seen so far scan indiscriminately, without regard to the potential vulnerable set.  However, worms are getting smarter and now seem to scan more efficiently -- concentrating on their potential prey \cite{moopaxsav03}.  Thus, the model is accurate in modeling smarter worms and provides an upper bound on the performance of worms that scan indiscriminately.

The fluid model is constructed as follows. Consider any one infected host.  The probability of its choosing a susceptible host for infection is $\frac{S(t)}{N}$.
Let the average time taken for infecting a susceptible host be $\frac{1}{\alpha}$.
%probability of infection is $\alpha$ (it could be $1$), the probability of infecting the chosen host is $\frac{\alpha S(t)}{N}$.
Then if the infected host chooses to scan $Q$ hosts in a unit of time, and there are $I(t)$ infected hosts performing the same kind of Bernoulli trials, then as $N\rightarrow \infty$ the expected number of infected hosts in a unit time is $Q \alpha\ I(t)\ S(t)/{N}$.

The factor $Q\alpha$ is the maximum number of susceptible hosts that an infected host can infect per unit time.  We define $\beta\triangleq Q \alpha$, which we call the \emph{virulence} of the worm.  In this paper we are primarily interested in the order relations of the system with $N$.  So we take the unit of time as the expected time taken for an infection $(1/\beta)$, which we call \emph{infection time units} (ITU).  Note that we may convert ITU to actual time by just multiplying by this factor.  Then with time measured in ITU, the expected number of infected hosts in an ITU is
 \BEQA
\lambda\triangleq \frac{I(t)\ S(t)}{N},
\EEQA
 with $\lambda$ being the \emph{rate} of infection .

Now, we assume that the infection process is Markovian, with time taken for infection to be exponentially distributed with transition rate equal to $\lambda$, then it can be shown \cite{kurtz78} that as $N\rightarrow \infty$, the fraction of infected hosts $i(t)\triangleq \frac{I(t)}{N}$ converges to
\BEQA
i(t)=i(0)+\int_0^t\left(1-i(s)\right)i(s)\ ds,
\EEQA
where we have used (\ref{eqn_tot_hosts}). We represent the above showing explicit dependence on $N$ (in differential form) as
\BEQA\label{eqn_worm_dissem}
\frac{dI(t)}{dt}=\frac{\ S(t)I(t)}{N},
\EEQA
where it is understood that $N$ is large.

The above is identical to the classical simple epidemic model \cite{fra80} and has been used successfully in modeling the spread of infectious diseases.  It has the closed form solution
\BEQA\label{eqn_sigmoid}
I(t)= \frac{I(0) e^{ t}}{1-\frac{I(0)}{N}\left(1-e^{ t}\right)}.
\EEQA
The plot of the above expression looks much like Figure \ref{fig_caida_worm} and it grows exponentially initially and then levels off, yielding the classic sigmoidal shape.

\subsection*{How long does it take for the worm to spread to a large number of hosts?}

Given that worms so far either follow a spread-then-attack mode of operation or cause gradual damage, it would be interesting to know the order of time by which a large number of hosts are infected.  We could possibly expect an attack (or significant damage) to occur at this time.  It also gives a rough benchmark time at which we can compare the performance of different patching schemes.  We use the following notation that defines a set of functions $\Theta(g(N))$.  We say $f(N)\in \Theta(g(N))$ if $\exists$ $c_1$, $c_2$ and $M$ such that
\BEQA
c_1\ g(N)\leq f(N)\leq c_2\ g(N)\quad \forall N\geq M
\EEQA
% \BEQA
% \lim_{N\rightarrow\infty}c\ \frac{f(N)}{g(N)}= 1,\nonumber
% \EEQA
% where $c$ is constant.

\begin{theorem}  The time by which significant spread of the worm occurs is $\Theta(\ln N)$.
\end{theorem}
\begin{proof}
 We would like to know when $I(t)=\kappa\ N$, where $0<\kappa<1$.  From (\ref{eqn_sigmoid}), we directly have
\BEQA
&~&I(0)e^{ t}=\kappa\ N-\kappa I(0)\left(1-e^{ t}\right)\nonumber\\
%&\Rightarrow& I(0)(1-\kappa)e^{ t}=\kappa(N-I(0))\nonumber\\
%&\Rightarrow& e^{ t}=\frac{\kappa(N-I(0))}{I(0)(1-\kappa)}\nonumber\\
%&\Rightarrow& t=  \ln\left(\frac{\kappa(N-I(0))}{I(0)(1-\kappa)}\right)\nonumber\\
&\Rightarrow& t=  \ln \frac{\kappa}{1-\kappa} + \ln \frac{N-I(0)}{I(0)}\nonumber
\EEQA
For fixed $\kappa$ this time is $\Theta(\ln N)$.
\end{proof}

The above result says that the worm spreads exponentially fast in any relevant time-frame. We consider an example to illustrate what this means.\\
%\vspace{0.2in}

\noindent \textbf{Example}

Consider a worm with a virulence of $\beta=1.5$ hosts per minute and a susceptible population of $85,000$ hosts.  It would take $\ln N= 11.3$ ITU or  about $8$ minutes to infect a significant population.  The performance of such a worm is comparable to that of a worm like Slammer \cite{moopaxsav03} that spread to $75,000$ hosts in $10$ minutes.  Other worms have much lower rates of spread due to poor design of the scanning mechanism. \hfill {\large $\blacksquare$}

The result also characterizes the time available for countermeasures once the worm has appeared. Countermeasures are useful only if they can do something about the problem in $\Theta(\ln N)$ time, otherwise it is a futile activity. We will keep this in mind while studying patching schemes.
%Thus, the efficacy of a patching mechanism may be understood by checking the number of infected hosts at $\Theta(\ln N)$ time.  We will use this as a metric while studying patch dissemination.

% This explains the rapid spread of smarter worms like Slammer which targets only susceptible hosts.  It also means that the time available for
%For example, the time constant of a system is usually considered to be the time taken for a $95\%$ change in state and

\section{Patch Dissemination}\label{patch_dissemination}

The propagation of worms can be halted by fixing the holes in the application that allows them to do so.  This is the point of patching.  As mentioned in the introduction, in most instances so far a patch has been developed sufficiently quickly that the number of infected hosts at the time that the patch is released is small, so active defense by patching is possible \cite{lilnic04}.  Hosts must be informed about the availability of the patch, which we assume takes a short time since it is a simple update message.
We are then faced with the second task of ensuring that all hosts obtain the patch.  %As mentioned in the introduction we assume that the worm does not impair the ability of the system to patch itself, i.e., it does not kill the host before it has spread to a sufficiently large number of hosts, which is in accordance with worms seen thus far.
Once patched, a host that was infected cannot be reinfected by the worm.  So the patching process reduces both the susceptible and the infected population, and eventually the system is worm free.  We then have the following metrics to characterize any particular method of patching:
%Intuitively it seems obvious that the number the graph of the number of infected hosts would be unimodal, i.e., it would increase, hit a maximum, and then decrease to zero.  Also, as we have seen earlier, the worm might possibly cause significant damage at $\Theta(\ln N)$ time.  The observations suggest that there are three major metrics from which the performance of a patch dissemination method could be compared:
\begin{itemize}
%\item How long does it take for the infection to start decreasing?
\item When does the infection hit is peak, and what is the number of infected hosts at this time?
\item How long does it take to end the infection?
\end{itemize}
 We must answer the above questions keeping in mind the fact that the worm might possibly cause significant damage at $\Theta(\ln N)$ time.  Our emphasis will be on the order relations in the system.
%The  patch dissemination mechanism has this much time in which to disburse the patch as widely as possible.
We will study two possible methods of patch dissemination:
\begin{enumerate}
\item A system with a fixed number of patch servers.
\item A peer-to-peer network.
\end{enumerate}
The system with a  fixed number of patch servers models either a dedicated bank of patch servers or that of a content distribution network (CDN) with a fixed number of replicas, while the P2P system models either a patching worm or a CDN that is implemented in a P2P fashion.

\subsection*{Fixed Number of Patch Servers}

Suppose the creator of the patch has a fixed number of patch servers.  Both infected and susceptible hosts try to download patches from the patch servers.
So the question arises whether a fixed number of servers can contain the spread of the worm.  %This method of patching is believed to be the one used in handling the Code-Red worm \cite{lilyua02}.
Let the number of servers be  $\overline{P}$, which is much smaller than the total number of hosts present in the network.  Let each server be capable of disseminating $\gamma$ patches in an ITU.  In other words, the actual maximum rate at which each server can disburse patches is $\frac{\gamma}{\beta}$ patches per unit time.  Then the rate at which the servers patches get disseminated is $\gamma \overline{P}$ patches per ITU, until the number of hosts to be patched is less that $\overline{P}$.  After this point the rate is equal to the number of hosts remaining times $\gamma$.  This finishing phase is irrelevant to our study, since the number of hosts patched during this time is just $\overline{P}$.  We now construct the fluid differential equations corresponding to the system.

Let number of patched hosts at time $t$ be denoted by $P(t)$.  As before, the number of infected and susceptible hosts at this time are $I(t)$ and $S(t)$ respectively.  Also, the rate at which the worm grows is $S(t)\ I(t)/N$.  However, patching causes the number of infectious hosts in the network to decrease.   Servers disburse patches to both infected as well as susceptible hosts.  Then the expected number of infected hosts that obtain the patch in a unit time is $(\gamma\ \overline{P}\ I(t))/(S(t)+I(t))$.  In the fluid model, this quantity is the rate at which the infection decreases.  Similarly, the rate at which susceptible population decreases is $(\gamma\ \overline{P}\ S(t))/(S(t)+I(t))$.  However, since the total number of hosts in the system is fixed, we can describe the system in terms of the infected and patched hosts alone as follows:
\BEQA
\frac{d\ P(t)}{dt}&=&\gamma\ \overline{P}\label{eqn_p1_patch}\\
\frac{d\ I(t)}{dt}&=&\frac{S(t)\ I(t)}{N}-\frac{\gamma\ \overline{P}\ I(t)}{S(t)+I(t)}\label{eqn_p1_infect}\\
N&=&S(t)+I(t)+P(t)\label{eqn_p1_total}
\EEQA
The differential equations are valid when number of patched hosts is no greater than $N-\overline{P}$, which is practically till all the hosts are patched since $N>>\overline{P}$.  We then have the following theorem:
%%%%%%%%%%
%%%%%%%%%%
\begin{theorem}\label{them_p1}
For the fixed number of servers paradigm, we have that the number of infected hosts
\BEQA
I(t)=\frac{\left(N-\overline{P}-\gamma\ \overline{P}\ t\right)\left(\exp \left(t-\frac{\overline{P}\ t}{N}-\frac{\gamma\ \overline{P}\ t^2}{2N}\right)\right)}{\exp \left(t-\frac{\overline{P}\ t}{N}-\frac{\gamma\ \overline{P}\ t^2}{2N}\right) + C},\label{eqn_p1_infections}
\EEQA
where $C=(N-\overline{P})/I(0)\in\Theta(N)$  and $t\in[0,\frac{N-2\overline{P}}{\gamma\overline{P}}]$.
%$\Theta(\ln N)$ time is $\Theta(N)$.
\end{theorem}
\begin{proof}
From (\ref{eqn_p1_patch}) by simple integration, with the initial condition $P(0)=\overline{P}$, we have
\BEQA
P(t)%&=&\int_0^t \gamma\ \overline{P}\ ds + constant\nonumber\\
=\gamma\ \overline{P}\ t + \overline{P}\label{eqn_p1_int_patch}
\EEQA
 %So the number of patched systems increases linearly with time.
So the time at which the number of patched hosts is $N-\overline{P}$ is $t=\frac{1}{\gamma \overline{P}}\left(N-2\overline{P}\right)$.  Now, consider the infection process.  From (\ref{eqn_p1_infect}) and (\ref{eqn_p1_total}) we have
\BEQA
\frac{d\ I(t)}{dt}%&=& \frac{S(t))\ I(t)}{N}-\frac{\gamma\ \overline{P}\ I(t)}{N-P(t)}\\
&=& \left(\frac{N-P(t)-I(t)}{N}\right)I(t)-\frac{\gamma\ \overline{P}\ I(t)}{N-P(t)}\nonumber\\
&=& -\frac{I^2(t)}{N} + \left(1-\frac{{P(t)}}{N}-\frac{\gamma\ \overline{P}}{N-P(t)}\right)I(t)\nonumber
\EEQA
Rearranging the above, we have the following second order Bernoulli differential equation
\BEQA
\frac{d\ I(t)}{dt} - \left(1-\frac{{P(t)}}{N}-\frac{\gamma\ \overline{P}}{N-P(t)}\right)I(t)= -\frac{I^2(t)}{N}\nonumber
\EEQA
Substituting $V(t)=\frac{1}{I(t)}$ yields a first order differential equation of form
\BEQA\label{eqn_p1_diffneqn}
\frac{d\ V(t)}{dt} + \left(1-\frac{{P(t)}}{N}-\frac{\gamma\ \overline{P}}{N-P(t)}\right)V(t)= \frac{1}{N}
\EEQA

The solution to (\ref{eqn_p1_diffneqn}) is of the form
\BEQA\label{eqn_p1_solnform}
V(t)=\frac{\frac{1}{N}\int J(t)\ dt + C}{J(t)},
\EEQA
where $C$ is a constant and
\BEQA
%J(t)=\nonumber\\
J(t)=\exp \left(\int\left(1-\frac{P(t)}{N}-\frac{\gamma\ \overline{P}}{N-P(t)}\right)\ dt\right)\hspace{0.2in}\nonumber\\
=\left(N-\overline{P}-\gamma\ \overline{P}\ t\right) \exp \left(t-\frac{\overline{P}\ t}{N} -\frac{\gamma\ \overline{P}\ t^2}{2N}\right)\label{eqn_p1_intfctr}
\EEQA
Here we have used the expression for $P(t)$ from (\ref{eqn_p1_int_patch}).  We now need to evaluate $\frac{1}{N}\int J(t)\ dt$.  This is accomplished by simple integration using the expression for $J(t)$ from (\ref{eqn_p1_intfctr}) as follows:
\BEQA
\frac{1}{N}\int J(t)\ dt=\hspace{2in}
\nonumber\\\int \left(1-\frac{\overline{P}}{N}-\frac{\gamma\ \overline{P}\ t}{N}\right) \exp \left(t-\frac{\overline{P}\ t}{N} -\frac{\gamma\ \overline{P}\ t^2}{2N}\right)\ dt\nonumber.
\EEQA
Making the substitution $q=t-\frac{\overline{P}\ t}{N} -\frac{\gamma\ \overline{P}\ t^2}{2N}$, and integrating we obtain
\BEQA
 \frac{1}{N}\int J(t)\ dt &=& \int \ e^{q}\ dq\nonumber\\
&=& \exp \left(t-\frac{\overline{P}\ t}{N}-\frac{\gamma\ \overline{P}\ t^2}{2N}\right)\label{eqn_p1_intj}
\EEQA

Thus,   (\ref{eqn_p1_solnform}), (\ref{eqn_p1_intfctr}) and (\ref{eqn_p1_intj}) yield the final answer
\BEQA
V(t)=\frac{\exp \left(t-\frac{\overline{P}\ t}{N}-\frac{\gamma\ \overline{P}\ t^2}{2N}\right) + C}{\left(N-\overline{P}-\gamma\ \overline{P}\ t\right)\left(\exp \left(t-\frac{\overline{P}\ t}{N}-\frac{\gamma\ \overline{P}\ t^2}{2N}\right)\right)}\label{eqn_p1_final}
\EEQA
Note that $C=(N-\overline{P})/I(0)\in \Theta(N)$, as seen by plugging in $t=0$.  Noting that $I(t)=1/V(t)$, we have the proof.
%
%Now, we would like to know what $V(t)$ looks like at $\Theta(\ln N)$ time, and from this calculate $I(\ln N)$.
%Substituting $t=c\ln N$ in  (\ref{eqn_p1_final}), we see that $V(\Theta(\ln N))$ is $\Theta(1/N)$, so $I(\Theta(\ln N))$ is $\Theta(N)$.
\end{proof}

We see how similar the expression for $I(t)$ looks to (\ref{eqn_sigmoid}).  %Except for the terms that are linear in $t$, it is practically identical.
Essentially, the infection progresses unhindered for small $t$.  We expect that the effect of patching will not be felt till a fairly large number of hosts is infected. We are now  ready to answer questions regarding its performance. We would first like to know when the number of infected hosts hits its maximum value.

\begin{corollary}\label{cor_p1_1}
For the fixed number of servers paradigm, the number of infected hosts is unimodal and starts decreasing when $t=2\ln\left(\frac{N}{\sqrt{\gamma\overline{P}I(0)}}\right)\in\Theta(\ln N)$.
\end{corollary}
\begin{proof}
To find out when the number of infected hosts starts decreasing, we need to find the time when $\frac{d\ I}{dt}\leq 0$. In order to do this we differentiate (\ref{eqn_p1_infections}) and obtain
\BEQA
\frac{d\ I}{dt} =\hspace{2.6in}\nonumber\\
\frac{e^X}{\left(C+ e^X\right)^2} \left(-\gamma\overline{P} \left( C + e^X\right)
+ \frac{C}{N} (N-\overline{P}-\gamma\overline{P} t)^2\right),\label{eqn_p1_derivative}
\EEQA
where
\BEQA
X\triangleq t-\frac{\overline{P}\ t}{N}-\frac{\gamma\ \overline{P}\ t^2}{2N}\nonumber
\EEQA
Setting $\frac{d\ I}{dt}\leq0$, substituting the value of $C$, and rearranging, we get
\BEQA
M(t)\triangleq\frac{\frac{N-\overline{P}}{I(0)N}\left(N-\overline{P}-\gamma\overline{P}t\right)^2}{\gamma\overline{P}\left(\frac{N-\overline{P}}{I(0)} +e^X\right)}\leq 1
\EEQA

We observe that for $t\geq 2\ln\left(\frac{N}{\sqrt{\gamma\overline{P}I(0)}}\right)$, we have that $M(t)\leq 1$ for large $N$.
%$M(t)\rightarrow k$, where $k\leq 1$ as $N\rightarrow\infty$.
Thus, for $t\in \Theta(\ln N)$,
%the two terms in (\ref{eqn_p1_derivative}) are both of order $N^2$.  So, for appropriate constants,
the number of infected hosts starts decreasing.
\end{proof}

Recall that in the system without patching, the time taken for infection of a significant population is $\Theta(\ln N)$.  So, the effect of patching is felt at exactly this time frame.  It also shows increasing the patching rate $\gamma \overline{P}$ has little effect unless it is impractically large (comparable to $N$).  So even if the patch servers work very fast as compared to the virus, there would be no major consequence on the time at which the infection decreases.  We next consider the question of how many hosts are infected at this time.  Because the graph is unimodal, this is also the time at which the maximum number of hosts is infected.

\begin{corollary}\label{cor_p1_2}
For the fixed number of servers paradigm, the number of infected hosts is $\Theta(N)$ for
%$t=2\ln\left(\frac{N}{\sqrt{\gamma\overline{P}I(0)}}\right)$
$t\in \Theta(\ln N)$.  This is also the maximum number of infected hosts over all time.
\end{corollary}
\begin{proof}
Consider (\ref{eqn_p1_infections}).  For $t\in \Theta(\ln N)$, the number of infected hosts is $\Theta(N)$.
\end{proof}

The above result implies that a fixed number of patch servers is simply unable to cope with the spread of a well designed worm!  In an unpatched system, the worm spreads to $\Theta(N)$ hosts in $\Theta(\ln N)$ time.  Thus, as far as the worm is concerned, a system with a fixed number of patch servers behaves as if practically no patching were occurring up to $\Theta(\ln N)$ time.  A worm which timed its attack at $\Theta(\ln N)$ time would be unstoppable. The next question is that of when the infection actually dies down, i.e., how long will it take for the number of infected hosts to come down to $\Theta(1)$?

\begin{corollary}\label{cor_p1_3}
For the fixed number of servers paradigm, the time taken for the number of infected hosts to decrease to $\Theta(1)$ is $t=\frac{N-2\overline{P}}{\gamma\overline{P}}\in \Theta(N)$.
\end{corollary}
\begin{proof}
From (\ref{eqn_p1_int_patch}), substituting $t=\frac{N-2\overline{P}}{\gamma\overline{P}}$, we have that $\lim_{N\rightarrow \infty} I(t)=\overline{P}\in\Theta(1)$.  Hence the proof.
\end{proof}

Thus, the infection is contained well after the attack takes place.  We conclude that patching with a fixed number of servers is a futile activity.  Clearly, we don't just need a patch that kills the worm on contact, but also an efficient distribution mechanism that can deal with the worm by creating new servers -- a P2P system.

\subsection*{Peer-to-Peer Patch Dissemination}

We have just seen that the fixed number of patch servers scheme performs extremely badly in disseminating patches.  We would like to design a system that matches the worm in its capability to proliferate.  The obvious solution is to use a P2P model.  A patch received from a peer would have to checked with respect to a hash (sent with the update message, for instance) to ensure security of patches.  Such a method of verification has already been implemented in BitTorrent \cite{coh03}.  In the proposed scheme, hosts use a \emph{pull} mechanism to obtain the patch, i.e., they contact hosts at random and ask them if they have the patch.  If the patch is available, it is downloaded, verified, and installed. This mechanism is at variance with the \emph{push} structure of the worm, in which infected hosts contact hosts at random and try to infect them.  However, there is no real difference in the fluid model.%We assume that for the time duration that we are interested, the system does not have arrivals and departures of hosts.  The model is valid for the case of flash crowds all trying to get the same file in a small time interval, which is what we expect in our scenario

Construction of the fluid model is similar to what we have seen before.  Let the number of hosts that initially possess the patch be  $\overline{P}$, which is much smaller than the total number of hosts present in the network. Let each host be capable of disseminating a maximum of $\gamma$ patches in an ITU.  Note that $\gamma$ is likely to be smaller than the $\gamma$ that we encountered in the fixed number of servers case, since the hosts in a P2P system are not dedicated patch servers. %However, with appropriate throttling, $\gamma$ can be made close to $1$, i.e., the expected time for an infection and a patch could be made similar.
The rate of  patch dissemination looks very similar to the rate of worm dissemination that we saw in (\ref{eqn_worm_dissem})  and is given by $\frac{\gamma}{N}\ (S(t) + I(t))\ P(t)$.  Also, while the rate at which the worm increases is still $\frac{1}{N}\ S(t)\ I(t)$, it now decreases at the rate at which infected hosts are patched, which is just $\frac{\gamma}{N}\ I(t)\ P(t)$.  Then we have the following description of the system:
\BEQA
\frac{d\ P(t)}{dt}&=&\frac{\gamma}{N}\ (S(t) + I(t))\ P(t)\label{eqn_p2_patch}\\
\frac{d\ I(t)}{dt}&=&\frac{1}{N}S(t)\ I(t)-\frac{\gamma}{N}\ I(t)\ P(t)\label{eqn_p2_infect}\\
N&=&S(t)+I(t)+P(t)\label{eqn_p2_total}
\EEQA
Our problem is now to solve the above system of equations and answer questions regarding the performance of the scheme.
%%%%%%%%%%%%%%
%%%%%%%%%%%%%%
\begin{theorem}\label{them_p2_1}
For the P2P paradigm, the number of infected hosts at time $t$ is given by
%$\ln N$ time is $\Theta(1)$.
%, i.e., for $\gamma=1$ this number of infected hosts is $\Theta(\frac{1}{N})$.
\BEQA
I(t)=\hspace{2.5in}\nonumber\\
\left( \frac{1}{N^2}\frac{\overline{P} e^{\gamma t}}{1-\frac{\overline{P}}{N}} + \frac{1}{N} \right.\hspace{1.5in}\nonumber\\
\left.+\  Ce^{\gamma t}\left( \frac{\overline{P}}{N} + \left(1-\frac{\overline{P}}{N}\right) e^{-\gamma t} \right) ^{\frac{1}{\gamma}+1}\right)^{-1},\label{eqn_p2}
\EEQA
where $C=1/I(0)\in\Theta(1)$ for large N.
\end{theorem}
\begin{proof}
The proof technique is similar to the one used earlier.  We first solve for $P(t)$ using (\ref{eqn_p2_patch}) and (\ref{eqn_p2_total}) which is known to have the solution (of the same form as (\ref{eqn_sigmoid}))
\BEQA
P(t)= \frac{\overline{P} e^{ \gamma t}}{1-\frac{\overline{P}}{N}\left(1-e^{\gamma t}\right)}\label{eqn_p2_sigmoid}
\EEQA

We then use (\ref{eqn_p2_infect}) and (\ref{eqn_p2_total}) to obtain
\BEQA
\frac{d\ I(t)}{dt}%&=& \frac{S(t))\ I(t)}{N}-\frac{\gamma\ \overline{P}\ I(t)}{N-P(t)}\\
&=& \left(\frac{N-P(t)-I(t)-\gamma\ P(t)}{N}\right)I(t)\nonumber\\
&=& -\frac{I^2(t)}{N} + \frac{1}{N}\left(N-(1+\gamma)P(t)\right)I(t)\nonumber
\EEQA
Rearranging, we have the following second order Bernoulli differential equation
\BEQA
\frac{d\ I(t)}{dt} - \left(1-\frac{1+\gamma}{N}\ P(t)\right)I(t)= -\frac{I^2(t)}{N}.\nonumber
\EEQA
%where $\zeta\triangleq (1+\gamma)/N$.
We convert the above into a first order differential equation by substituting $V(t)=\frac{1}{I(t)}$ and obtain
\BEQA\label{eqn_p2_diffneqn}
\frac{d\ V(t)}{dt} + \left(1-\frac{1+\gamma}{N} P(t)\right)V(t) = \frac{1}{N}
\EEQA

As before, the above equation has a closed form solution given by
\BEQA\label{eqn_p2_closed}
V(t)=\frac{\frac{1}{N}\int J(t)\ dt + C}{J(t)},
\EEQA
where $C$ is a constant and
\BEQA
J(t)&=&\exp \left(\int\left(1-\frac{1+\gamma}{N}\ P(t)\right)\ dt\right)\nonumber\\
&=& \exp \left(t-\frac{1+\gamma}{\gamma} \ln \left(1-\frac{\overline{P}}{{N}}\left(1-e^{\gamma t}\right)\right)\right)\nonumber\\
&=& \frac{e^{t}} {\left( 1-\frac{\overline{P}}{{N}}\left(1-e^{\gamma t}\right)\right)^{\frac{1+\gamma}{\gamma}}}\nonumber\\
&=&\frac{ e^{-\gamma t} } { \left( \frac{\overline{P}}{N} + \left(1-\frac{\overline{P}}{N}\right)e^{-\gamma t}\right)^{\frac{1+\gamma}{\gamma}}}
\label{eqn_p2_intfctr}
\EEQA
Here we have used the expression for $P(t)$ from (\ref{eqn_p2_sigmoid}).  Now, in order to obtain the closed form solution, we also require $\frac{1}{N}\int J(t)\ dt$.  So we proceed to integrate the above expression.  We have
\BEQA
\frac{1}{N}\int J(t)\ dt=\frac{1}{N}\int \frac{ e^{-\gamma t} } { \left( \frac{\overline{P}}{N} + \left(1-\frac{\overline{P}}{N}\right)e^{-\gamma t}\right)^{\frac{1+\gamma}{\gamma}}}\ dt\nonumber
\EEQA
We make the substitution $q=e^{-\gamma t}$ and obtain the relation
\BEQA
\frac{1}{N}\int J(t)\ dt= \frac{-1}{\gamma N}\int \frac{dq}{\left( \frac{\overline{P}}{N} + \left(1-\frac{\overline{P}}{N}\right)q\right)^{\frac{1+\gamma}{\gamma}}}\ dt\nonumber\\
= \frac{1}{N\left(1-\frac{\overline{P}}{N}\right)}\left( \frac{\overline{P}}{N} + \left(1-\frac{\overline{P}}{N}\right)q\right)^{\frac{-1}{\gamma}}\hspace{0.1in}\nonumber\\
= \frac{1}{N\left(1-\frac{\overline{P}}{N}\right)}\left( \frac{\overline{P}}{N} + \left(1-\frac{\overline{P}}{N}\right)e^{-\gamma t}\right)^{\frac{-1}{\gamma}}\label{eqn_p2_intj}
\EEQA

 Then using (\ref{eqn_p2_closed}), (\ref{eqn_p2_intfctr}) and (\ref{eqn_p2_intj}), and simplifying we obtain
\BEQA
V(t)= \hspace{2.8in}\nonumber\\
\frac{1}{N^2}\frac{\overline{P} e^{\gamma t}}{1-\frac{\overline{P}}{N}} + \frac{1}{N} + Ce^{\gamma t}\left( \frac{\overline{P}}{N} + \left(1-\frac{\overline{P}}{N}\right) e^{-\gamma t} \right) ^{\frac{1}{\gamma}+1}\label{eqn_p2_final}
\EEQA
Note that $C=1/I(0)\in\Theta(1)$ for large $N$, as seen by plugging in $t=0$.  Finally using the fact that $I(t)=1/V(t)$ (by definition) we have the proof.
\end{proof}
% with a positive value of $V(0)=\frac{1}{I(0)}$.
%
%Finally, we evaluate the above at $t=c\ln N$.  We see that for $c=2/\gamma$, $V(c\ln(N))$ is of order $1$.  Thus, $I(\Theta(\ln N))\in \Theta(1)$.

The result shows that as expected, the patch spreads exponentially, directly competing with and destroying the worm. We can perform a similar analysis as we did in the fixed number of servers case to determine when the infection starts decreasing.  We have the following result:

\begin{corollary}\label{cor_p2_1}
For the P2P paradigm, the number of infected hosts is unimodal and decreases for $t\geq\frac{1}{\gamma}\ln \left(\frac{N}{{\gamma \overline{P}}}\right)\in\Theta(\ln N)$.
\end{corollary}
\begin{proof}
As before, the proof is obtained by differentiation.  Note that $V(t)=\frac{1}{I(t)}$, hence
\BEQA
\frac{d\ V(t)}{dt}=\frac{-1}{I^2(t)}\frac{d\ I(t)}{dt}, \nonumber
\EEQA
which means that we need to find the time at which $V(t)$ starts increasing.  Differentiating (\ref{eqn_p2_final}), and setting $\frac{d\ V(t)}{dt}\geq 0$, we obtain
\BEQA
\frac{\gamma \overline{P} e^{\gamma t}}{N^2\left(1-\frac{\overline{P}}{N}\right)} + \frac{\gamma}{I(0)}\left(\frac{\overline{P}}{N}+\left(1-\frac{\overline{P}}{N}\right)e^{-\gamma t}\right)^{\frac{1}{\gamma}+1}e^{\gamma t}\hspace{0.3in}\nonumber\\
 - \frac{\gamma}{I(0)} \left( \frac{1}{\gamma}+1\right)\left(1-\frac{\overline{P}}{N}\right)\left(\frac{\overline{P}}{N}+\left(1-\frac{\overline{P}}{N}\right)e^{-\gamma t}\right)^{\frac{1}{\gamma}}\geq 0\label{eqn_cor_p2_1_1}
\EEQA

%Now, we are looking for the smallest time above which $V(t)$ is increasing.  We observe that the first term is positive and is small for $t\leq \frac{2}{\gamma} \ln N$, also $\frac{\overline{P}}{N}$ is small compared to $1$. Thus, for

Since the first term is positive, and  $\frac{\overline{P}}{N}$ is small compared to $1$, a sufficient condition for large $N$ is
\BEQA
e^{\gamma t}\left(\frac{\overline{P}}{N}+e^{-\gamma t}\right)^{\frac{1}{\gamma}+1} - \left( \frac{1}{\gamma}+1\right) \left(\frac{\overline{P}}{N}+e^{-\gamma t}\right)^{\frac{1}{\gamma}}\geq 0\nonumber\\
%\Rightarrow e^{\gamma t}\geq \frac{N}{\gamma\overline{P}}\hspace{2.5in}\nonumber\\
\Rightarrow t\geq \frac{1}{\gamma} \ln \left(\frac{N}{{\gamma \overline{P}}}\right)\hspace{2in}\nonumber
\EEQA
%Note that even for $t\geq \frac{2}{\gamma} \ln N$, the inequality still holds.
%Letting $t\geq\frac{2}{1+\gamma}\ln \left(\sqrt{\frac{1+\frac{1}{\gamma}}{I(0)\overline{P}}}N\right)$ causes $\frac{d\ V(t)}{dt}\geq 0$ for large $N$.
%$\lim_{N\rightarrow\infty}\frac{d\ V(t)}{dt}=k$, where $k\leq 0$.
Note that the first term in (\ref{eqn_cor_p2_1_1}) is small for  $t\leq \frac{2}{\gamma} \ln N$. So the condition on $t$ is actually tight for large $N$.
Thus, for $t\geq\frac{1}{\gamma} \ln \left(\frac{N}{{\gamma \overline{P}}}\right)$, the number of infected hosts is decreasing.  Hence the proof.
\end{proof}

The result says that even in the P2P case, it would take $\Theta(\ln N)$ time for the infection to start decreasing. It also says that the time at which the infection starts decreasing is unaffected by the initial number of infected hosts, unlike the fixed server case.

The number of infected hosts at this time (which is also the maximum) ought to be much lower than in the fixed servers case since far more hosts have been patched in this time.  We show that this is indeed true in the following result:
\begin{corollary}\label{cor_p2_2}
For the P2P paradigm, the maximum number of infected hosts is
\BEQA
\Theta(N^{\frac{1}{\gamma}}) &\mbox{for}& \gamma > 1\nonumber\\
\Theta(N) &\mbox{for}&\gamma\leq1.\nonumber
\EEQA
\end{corollary}
\begin{proof}
The proof follows directly by substituting $t=\frac{1}{\gamma} \ln \left(\frac{N}{{\gamma \overline{P}}}\right)$
% $t=\frac{2}{1+\gamma}\ln \left(\sqrt{\frac{1+\frac{1}{\gamma}}{I(0)\overline{P}}}N\right)$
in (\ref{eqn_p2}) and letting $N\rightarrow \infty$.  The maximum number of infected hosts for $\gamma > 1$ is
\BEQA
I_{max}=\frac{\gamma I(0)N^{\frac{1}{\gamma}}}{\overline{P}^{\frac{1}{\gamma}}\left(1+{\gamma}\right)^{1+\frac{1}{\gamma}}} \in \Theta\left(N^{\frac{1}{\gamma}}\right). \nonumber
\EEQA
For $\gamma \leq 1$, we get from (\ref{eqn_p2}) that $I_{max}\in \Theta(N)$.
Hence the proof.
\end{proof}

The above results shows that even a P2P system has limited effect in $\Theta(\ln N)$ time if the patching constant $\gamma\leq 1$.  This seems intuitively correct -- since the virulence $\beta$ of the worm has been normalized to $1$, only if $\gamma> 1$  will we observe significant reduction in the maximum number of infected hosts.  The final question is that of when the infection is stamped out, i.e., how long does it take for the number of infected hosts to become small?
\begin{corollary}\label{cor_p2_3}
For the P2P paradigm, the time taken for the number of infected hosts to decrease to $\Theta(1)$ is $t=\frac{1}{\gamma}\left(1+\frac{1}{\gamma}\right) \ln N \in\Theta(\ln N)$.
\end{corollary}
\begin{proof}
The proof follows directly from substituting $t=\frac{1}{\gamma}\left(1+\frac{1}{\gamma}\right) \ln N$ in (\ref{eqn_p2}) and letting $N\rightarrow \infty$.
\end{proof}

Thus, the time at which the infection to start decreasing and the time at which it is wiped out are both $\Theta(\ln N)$.  \emph{Soon after the infection hits its peak, it also disappears}.  If a worm  were to time its attack at $\Theta(\ln N)$ time, it would only have a marginal impact on the network.
%From the proof we see that a good estimate of the time by which the infection loses potency is $$.
%If by an appropriate throttling method, we bring down the virulence of the worm such that the patch spreads with $\gamma\approx 2$, then the number of infected hosts at $t=\ln N$ time is $\Theta(1)$, which is tiny as compared to $N$.
%
% For $N$ large and $t=0$, the above simplifies to
% \BEQA
% \gamma \overline{P} - C\nonumber
% \EEQA
% Noting that $C\leq 0$ for $I(0)\geq 1, \overline{P}\geq 1$ yields the proof.

\subsection*{Discussion}

It is interesting to compare the different results we have with regard to the effect of patching constant $\gamma$ on the time at which the infection starts to decrease and the maximum number of infected hosts.

%ren Corollary \ref{cor_p1_1} and \ref{cor_p2_1}, with regard to the effect of patching constant $\gamma$ on the time at which the infection starts to decrease.

In Corollary \ref{cor_p1_1}, $\gamma$ appears only within the logarithm.  So only a $\gamma$ that is comparable with $N$ has any real effect.  %Thus, we observed in our Code-Red propagation example that $\gamma=276$ ($\gamma\overline{P}=6,900$) had little impact on the time at which the infection decreases.
On the other hand, in Corollary  \ref{cor_p2_1}, $\gamma$ appears both inside and outside the logarithm. Inside the logarithm, it would have to be quite large to have any visible effect.  However, since it appears outside and operates on $\ln N$ as well, the effect of even $\gamma=2$  is significant.

Again, in Corollary \ref{cor_p1_2} we noticed that for any $\gamma\in\Theta(1)$, the maximum number of infected hosts was $\Theta(N)$.  Increasing $\gamma$ has no effect unless $\gamma$ is of $\Theta(N)$, which is physically impossible. On the other hand in Corollary \ref{cor_p2_2}, even increasing $\gamma$ by a small amount results in \emph{order} differences in the maximum number of infected hosts.

So even a small rate of patching by the peers of a P2P network has far more impact than an enormous rate  of a fixed number of servers.  The results illustrate the profound impact that throttling the worm can have on the system -- for a fixed number of patch servers throttling is of limited value, but in a P2P system throttling gains are magnified enormously.
Thus, if we use the patch provider's $\overline{P}$ servers as \emph{seed} servers for distributed patch delivery in a P2P system, we can truly achieve outstanding performance -- we wipe out the infection exponentially fast!

\section{Experiments}\label{experiments}

We use data measured on the Internet along with simulations to illustrate the fact that our analytical results, which assumed large $N$, can be used to make fairly good predictions on reasonably large systems, and so mirror reality.  We consider a Code-Red v2 type worm with a virulence $\beta=1.8$ infections per hour \cite{stapax02} and a susceptible population of $360,000$ hosts (seen from Figure \ref{fig_caida_worm}).  The spread of this worm was measured in \cite{mooshabro02} and we obtained the data used in their study courtesy of CAIDA (www.caida.org).  Our simulations were performed by using Simulink to simulate the fluid differential equations.

\begin{figure}[htbp]
\begin{center}
\epsfig{file=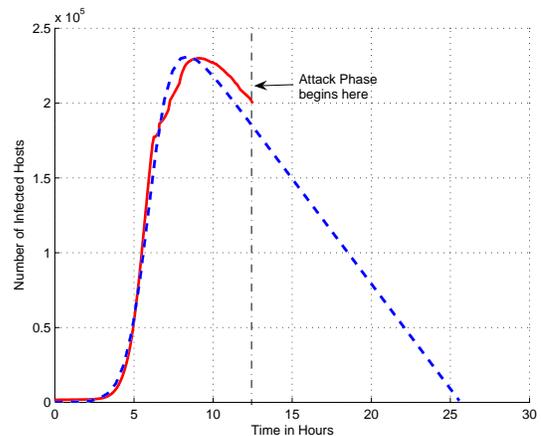,width=3.2in}
\caption{Graph illustrating the performance of a fixed number of patch servers.  The solid line is measured Internet data, while the dashed line corresponds to the fluid model.}
\label{fig_fixed_patch}
\end{center}
\end{figure}

We perform our  first experiment on the system with a fixed number of patch servers.  This was probably the method used in handling Code-Red v2, as the rate of patching with time was seen to be linear \cite{mooshabro02,lilyua02}, with about $15,000$ hosts being patched in $8$ hours.  The data obtained from Internet measurement is plotted as a solid line in Figure \ref{fig_fixed_patch}, while the dashed line is the simulation.   We calculated from the data that the patching rate $\gamma \overline{P}$ was roughly $7,800$ per ITU. We assume that $\overline{P}=25$ (this number is not important since only $\gamma \overline{P}$ has an effect on the system) and $I(0)=25$.   The zero for time was chosen by matching the exponential growth phase of the measured data with that of the simulation.  Measurement stopped when the worm went into attack mode and so stopped random scanning.
   %The same parameters were used earlier in our comparison with Internet measurement data, and we expect that the analytical results should be fairly accurate in the simulation setting as well.
From Corollary \ref{cor_p1_1} we expect the time at which the infection hits its peak is $t=2\ln\left(\frac{N}{\sqrt{\gamma\overline{P}I(0)}}\right)$ ITU, i.e., about $7.5$ hours, which matches fairly well with the graph.  We also expect from Corollary \ref{cor_p1_2} that the maximum number of infected hosts would be of order $10^5$, while the graph shows this value as about $2.3\times 10^5$.  Finally, we expect from Corollary \ref{cor_p1_3} that the infection is wiped out in $\frac{N-2\overline{P}}{\gamma\overline{P}}$ ITU, which is about $25$ hours.  There is no Internet data on this number (since measurement stopped during the attack phase), but the simulation result matches well with this value.

We note that a fluid model to explain the behavior of Code-Red v2 was studied in \cite{zougontow02} and a figure reminiscent of Figure  \ref{fig_fixed_patch} was presented there.  However, the model in \cite{zougontow02} uses an epidemic type patch dissemination, which is not directly related to the number of patch servers or the ratio of the patching rate to the worm propagation rate $\gamma$.  Our model is explicitly in terms of these physically measurable quantities.  Further, the results in \cite{zougontow02} are numerical solutions, while we obtain closed-form solutions which allow us to analytically predict the performance of different schemes.

\begin{figure}[htbp]
\begin{center}
\epsfig{file=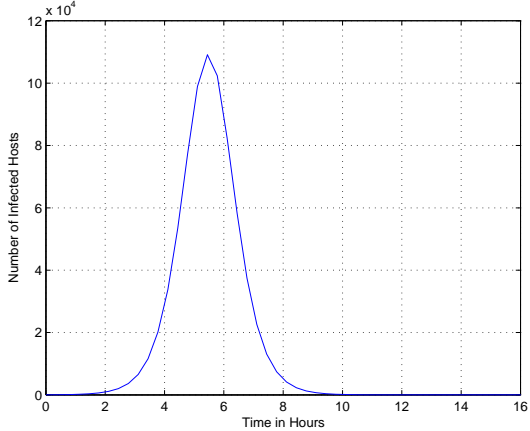,width=3.2in} \caption{Graph illustrating the performance of a P2P system with $\gamma=1$.}
\label{fig_p2p_patch_1}
\end{center}
\end{figure}

We next perform experiments with the P2P system.  Here we have no Internet data, since such a system has not been implemented.  However, we use simulations to illustrate that our order results are valid.  First we take $\gamma=1$, $\overline{P}=10$ and $I(0)=25$.  The results are shown in Figure \ref{fig_p2p_patch_1}.  We make use of the Corollaries \ref{cor_p2_1}, \ref{cor_p2_2} and \ref{cor_p2_3} to find the expected numerical values.  The expected time at which the infection starts reducing is $5.8$ hours, which matches well with the graph.  The number of infected hosts ought to be of the order $10^5$ at this time, and the graph shows a value of $1.1\times 10^5$.   Finally, the infection ought to end in about $14$ hours, which matches quite well with the simulation (the tail is difficult to see in the figure as the peak is quite high).  Notice that even with $\gamma=1$ the P2P system takes about half the time to wipe out the infection as the fixed server scheme.

\begin{figure}[htbp]
\begin{center}
\epsfig{file=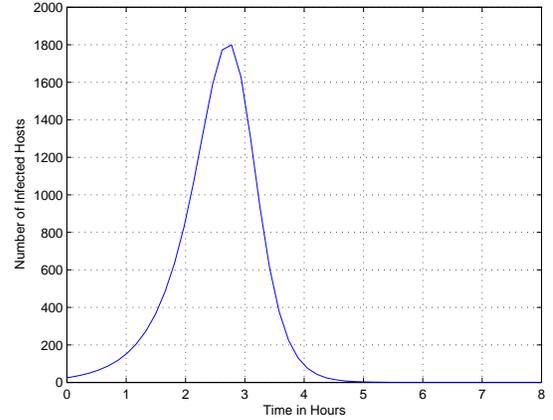,width=3.2in} \caption{Graph illustrating the performance of a P2P system with $\gamma=2$.}
\label{fig_p2p_patch_2}
\end{center}
\end{figure}

Our next experiment on the P2P system is to take $\gamma=2$, $\overline{P}=10$ and $I(0)=25$.  We wish to illustrate the effect of increasing $\gamma$ to $2$.  The results appear in  Figure \ref{fig_p2p_patch_2}.  The time at which we expect the infection to start decaying is $2.7$ hours, which is approximately what we see in the graph.  The number of hosts infected at this time should be of order $10^3$, which compares with $1.8\times 10^3$  that we see in the graph. Notice that both the time at which decay begins as well as the maximum number of infected hosts has shrunk sharply. The effect becomes more and more pronounced as $\gamma$ is increased.  Finally, we expect that the infection is over in $5.3$ hours, which is what we see in the graph.

The simulations backup our analytical results indicating the strength of P2P patching -- much lower number of infections and a much lower time in which the infection is contained.

\section{How many hosts have to be monitored?}\label{monitors}

So far we have studied the rate of propagation of worms and tried to understand the performance of possible patching schemes.
We now consider a related problem which a network operator would be interested in -- the number of hosts that require monitoring in order to obtain knowledge of the existence of the worm.  Monitoring in this fashion yields data on the distribution of infected hosts and efficacy of patching, for instance the data presented in Figures \ref{fig_caida_worm} and \ref{fig_fixed_patch}.  Thus, monitoring provides information on questions like when the infection began, where it originates, how many hosts are infected at any time and so on. %Potentially, it could also be used to identify sections of the address space that could be isolated (although isolation would only provide temporary relief, since even a small number of infected hosts outside the isolated space would grow exponentially).
Monitoring is often carried out by passively observing an unused portion of the Internet address space -- a so called ``Network Telescope'' \cite{mooshavoe04}.  Since the telescope should normally not receive any packets, scans directed at it often correspond to worm attacks.  Another possibility is for the network operator to passively monitor a number of real hosts in their address space so as to obtain information about anomalous behavior.  It has also been suggested that a P2P network could be used to identify anomalous behavior of hosts \cite{san03,coscro05}.
%
%Note that we cannot start developing the patch once the worm shows up -- the worm would have already spread to many hosts.  As in the case of  a network telescope \cite{mooshavoe04},
In all cases, we are interested in answering the following questions:
\begin{itemize}
\item How many hosts have to be monitored in order to find \emph{one} instance of worm presence in the network in a short time?
\item How many hosts have to be monitored in order to find out how many infected hosts are present at a given time?
\end{itemize}

To answer the first question, we need to understand the behavior of the worm just after it comes into its existence. We have the following result
\begin{theorem}\label{them_init_exp}
A worm spreads exponentially fast initially, regardless of the patching scheme.
\end{theorem}
\begin{proof}
The proof follows from (\ref{eqn_sigmoid}) and Theorems \ref{them_p1}, \ref{them_p2_1} by taking $t<<N$.
\end{proof}

%consider the fact that the worm spreads exponentially fast initially even if the system is being patched.
We know that the time taken for effects of patching to show up is $\Theta(\ln N)$ (Corollaries \ref{cor_p1_1}, \ref{cor_p2_1}).  We would like to find out about the worm before it spreads to too many hosts.  Since the worm spreads exponentially initially, if we would like to know about the worm when the number of infected hosts is $\Theta(\ln N)$, we have to ensure that the monitors pick up its presence in $\Theta(\ln \ln N)$ time.  The question is that of how many hosts to monitor to obtain this information.

Suppose that we monitor $M$ hosts.  Then the probability that a particular infected host chooses one of these monitored hosts is $\frac{M}{N}$.  The expected number of monitored hosts  that are scanned in an ITU by all infected hosts is given by $\frac{M I(t)}{N}$.  So in the fluid model (with the same assumptions as before), this value is the rate at which monitored systems are scanned.  Then we have
\BEQA\label{eqn_monitor}
\frac{d \overline{M}(t)}{dt}=\frac{ M I(t)}{N},
\EEQA
where $\overline{M}(t)$ is the total number of scans received by monitored hosts in the time interval $[0,t]$.  We now have the following result.
\begin{theorem} In order to detect the worm by $t\in\Theta(\ln \ln N)$, the number of hosts that have to be monitored is
$\Theta\left(\frac{N}{\ln N}\right)$.
\end{theorem}
\begin{proof}
The proof is obtained by straightforward integration of (\ref{eqn_monitor}).  We have
\BEQA
\overline{M}(t)&=&\int_0^t\frac{ M I(s)}{N}\ ds\nonumber\\
&=&\frac{ M}{N}\int_0^t \frac{I(0) e^{ s}}{1-\frac{I(0)}{N}\left(1-e^{ s}\right)}\ ds\nonumber\\
&=&\frac{ M}{N} N\ \ln\left(1-\frac{I(0)}{N}\left(1-e^{ t}\right)\right)\nonumber,
\EEQA
where we have used (\ref{eqn_sigmoid}) in the second step and $\overline{M}(0)=0$.  We could equivalently use $I(t)=I(0)e^t$ as Theorem \ref{them_init_exp} suggests.

Since we would like $\overline{M}(t)$ to be of order $1$ in order to detect the worm at some time $t$, we have from the above that
\BEQA
M=\frac{N}{N \ln\left(1-\frac{I(0)}{N}\left(1-e^{ t}\right)\right)}\nonumber.
\EEQA
Then if we are to detect the worm in  $\ln \ln N$ time, we need
\BEQA
M=\frac{N}{N \ln\left(1-\frac{I(0)}{N}\left(1-e^{\ln \ln N}\right)\right)}\nonumber,
\EEQA
which is easily verified to be $\Theta\left(\frac{N}{\ln N}\right)$.
\end{proof}

The result indicates that as the number of hosts in the system increases, the number of hosts to be monitored in order to obtain fast information about the worm is extremely large.

We now consider the second question.  Suppose that we want to know how many hosts are infected at a particular time. We know from the previous section that the amount of time required for patching to take effect in the fixed number of patch server case is $\Theta(\ln N)$. If we assume that most of the infected hosts take this long to be patched, then an infected host is active for approximately this amount of time.  Any one infected host would perform $t$ scans in $t$ time (remember that $\beta$ has been normalized to $1$).  So the number of scans received by the monitor by a single infected host in $\Theta(\ln N)$ time (assuming that the infected host is not patched in the interval) is $\frac{M}{N} \ln N$. If we set this number equal to $1$, i.e., our monitor receives $1$ scan from a particular infected host, we need
\BEQA
\frac{M}{N} \ln N=1\nonumber\\
\Rightarrow M=\frac{N}{\ln N},
\EEQA
which is identical to the previous result.  Thus, the thumb rule of monitoring $\frac{N}{\ln N}$ hosts would give a good estimate of the events in the network.
However, if a $P2P$ method of patching were used, since patching occurs exponentially fast, one would have to monitor events at a finer time scale.  If we assume that the number of infected hosts that were patched in time $\Theta(\ln \ln N)$ is small (it could be a maximum of $\Theta(\ln N)$ which is small), then the state of the system remains relatively constant in this tiny time interval.  Proceeding as before we get that the number of hosts to be monitored is now $\frac{N}{\ln \ln N}$, which is even higher.
  The implications of the above results are best illustrated by examples.\\

\noindent\textbf{Example}

Consider the Slammer-like worm example, where the total number of susceptible hosts is $85,000$ and the virulence $\beta=1.5$ infections per minute.  We saw earlier that a significant population was compromised in $8$ minutes. Also, $\ln \ln N = 2.42$ ITU, which is $1.6$ minutes.  Thus, if  we want to know about the worm's presence in $1.6$ minutes, we would have to monitor $85,000/\ln (85,000) \approx 7,500$ hosts, which is close to $\frac{1}{10}$th of the population.  The number of infected hosts at this time would be of the order $\ln N=11$ hosts.\\

\noindent\textbf{Example}

Consider the current Internet, which largely runs IPv4.  There are a total of $2^{32}$ addresses present in the system.  If we would like to know about the presence of a worm in $\ln \ln N =3$ ITU, we would have to monitor $2^{32}/\ln 2^{32}\approx 2^{27}$ addresses -- which is larger than a $/8$ prefix!  To understand the events in a P2P system, we need to monitor practically all the hosts.\hfill {\large $\blacksquare$}

Our conclusion is that one cannot hope to achieve monitoring of such a large fraction of the Internet without the active participation of either the network operators or the hosts themselves.  The numbers strongly support the establishment of a P2P monitoring system, with either the network operators sharing data on anomalous behavior or by having  peers look out for abnormal activity such as repeated \emph{syn}s from an arbitrary host, and report it to a central monitor that would keep records.  A system with monitoring based on the lines suggested in \cite{san03,coscro05} might be the best way to accomplish this goal.

\section{Conclusions}\label{conclusions}
In this paper, we have sought to make a convincing case for the use of P2P networks for tackling Internet worms.
We first studied the classical epidemic fluid model in order to understand the time scales of events.
Using analysis, measured data and simulations, we then showed that a fixed number of patch servers is incapable of handling an epidemic.  We also showed that a P2P system is far better suited to handle worm outbreaks, both in terms off the maximum number of infected hosts,as well as the time taken to wipe out the infection.  Finally, we considered the issue of monitoring the network and showed that the number of monitors required is so large that one would need cooperation either among network operators or the hosts in the system to obtain reliable estimates.

We would like to extend our work to include a complete stochastic analysis of worms so as to place the fluid models on a sound mathematical foundation.  We would also like to understand the effects of more complex worm models in order to include second order effects.

\bibliographystyle{IEEEtran}
\bibliography{p2p_security}

\begin{thebibliography}{10}
\providecommand{\url}[1]{#1}
\csname url@rmstyle\endcsname
\providecommand{\newblock}{\relax}
\providecommand{\bibinfo}[2]{#2}
\providecommand\BIBentrySTDinterwordspacing{\spaceskip=0pt\relax}
\providecommand\BIBentryALTinterwordstretchfactor{4}
\providecommand\BIBentryALTinterwordspacing{\spaceskip=\fontdimen2\font plus
\BIBentryALTinterwordstretchfactor\fontdimen3\font minus
  \fontdimen4\font\relax}
\providecommand\BIBforeignlanguage[2]{{%
\expandafter\ifx\csname l@#1\endcsname\relax
\typeout{** WARNING: IEEEtran.bst: No hyphenation pattern has been}%
\typeout{** loaded for the language `#1'. Using the pattern for}%
\typeout{** the default language instead.}%
\else
\language=\csname l@#1\endcsname
\fi
#2}}

\bibitem{mooshabro02}
D.~Moore, C.~Shannon, and J.~Brown, ``{Code-Red: a case study on the spread and
  victims of an Internet worm},'' in \emph{Proceedings of Internet Measurement
  Workshop (IMW)}, Marseille, France, November 2002.

\bibitem{baicoojah05}
M.~Bailey, E.~Cooke, F.~Jahanian, and D.~Watson, ``{The Blaster worm: Then and
  Now},'' \emph{IEEE Security and Privacy Magazine}, vol.~3, no.~4, pp. 26--31,
  July 2005.

\bibitem{weaell04}
N.~Weaver and D.~Ellis, ``{Reflections on the Witty Worm: Analyzing the
  Attacker},'' \emph{;login: The USENIX Magazine}, vol.~29, no.~3, pp. 34--37,
  June 2004.

\bibitem{moopaxsav03}
D.~Moore, V.~Paxson, S.~Savage, C.~Shannon, S.~Staniford, and N.~Weaver,
  ``{Inside the Slammer worm},'' \emph{IEEE Security and Privacy Magazine},
  vol.~1, no.~4, pp. 33--39, July 2003.

\bibitem{lilnic04}
M.~Liljenstam and D.~Nicol, ``{Comparing passive and active worm defenses},''
  in \emph{Proceedings of the First International Conference on Quantitative
  Evaluation of Systems (QEST)}, Enschede, Netherlands, September 2004.

\bibitem{zougontowgao05}
C.~C. Zou, W.~Gong, D.~Towsley, and L.~Gao, ``{The monitoring and early
  detection of Internet worms},'' \emph{IEEE/ACM Transactions on Networking},
  vol.~13, no.~5, pp. 961--974, October 2005.

\bibitem{mooshavoe04}
D.~Moore, C.~Shannon, G.~Voelker, and S.~Savage, ``Network telescopes:
  Technical report,'' 2004, {Cooperative Association for Internet Data Analysis
  (CAIDA) Technical Report}.

\bibitem{san03}
J.~Sandin, ``{P2P systems for worm detection},'' in \emph{DIMACS Workshop on
  large scale attacks}, Piscataway, NJ,USA, September 2003.

\bibitem{coscro05}
M.~Costa, J.~Crowcroft, M.~Castro, A.~Rowstron, L.~Zhou, L.~Zhang, and
  P.~Barham, ``{Vigilante: End-to-End Containment of Internet Worms},'' in
  \emph{Proceedings of the 20th ACM Symposium on Operating Systems Principles
  (SOSP '05)}, Brighton, United Kingdom, October 2005.

\bibitem{kurtz78}
T.~G. Kurtz, ``Strong approximation theorems for density dependent markov
  chains,'' \emph{Stochastic Processes and their Applications}, vol.~6, pp.
  223--240, 1978.

\bibitem{fra80}
J.~C. Frauenthal, \emph{Mathematical Modeling in Epidemiology}.\hskip 1em plus
  0.5em minus 0.4em\relax Springer-Verlag, Berlin, Germany, 1980.

\bibitem{dalgan99}
D.~J. Daley and J.~Gani, \emph{{Epidemic Modelling: An Introduction}}.\hskip
  1em plus 0.5em minus 0.4em\relax Canbridge University Press, Cambridge, UK,
  1999.

\bibitem{zougontow02}
C.~C. Zou, W.~Gong, and D.~Towsley, ``{Code Red Worm Propagation Modeling and
  Analysis},'' in \emph{9th ACM Conference on Computer and Communication
  Security (CCS'02)}, Washington DC, USA, November 2002.

\bibitem{stapax02}
S.~Staniford, V.~Paxson, and N.~Weaver, ``{How to 0wn the Internet in Your
  Spare Time},'' in \emph{Proceedings of the 11th USENIX Security Symposium
  (Security '02)}, San Francisco, CA, USA, August 2002.

\bibitem{chegaokwa03}
Z.~Chen, L.~Gao, and K.~Kwiat, ``{Modeling the spread of active worms},'' in
  \emph{Proceedings of {IEEE} {INFOCOM} 2003}, San Franciso, CA, USA, April
  2003.

\bibitem{weaham04}
N.~Weaver, I.~Hamadeh, G.~Kesidis, and V.~Paxson, ``Preliminary results using
  scale-down to explore worm dynamics,'' in \emph{Proceedings of the Second
  Workshop on Rapid Malcode ACM-SIGSAC WORM 2005}, Washington, DC, USA, October
  2004.

\bibitem{keshamjiw05}
G.~Kesidis, I.~Hamadeh, and S.~Jiwasurat, ``{Coupled Kermack-McKendrick model
  for randomly scanning worms },'' in \emph{Proceedings of QoS-IP}, Sicily,
  Italy, February 2005.

\bibitem{vojgan05a}
M.~Vojnovi\'{c} and A.~J. Ganesh, ``On the effectiveness of automatic
  patching,'' in \emph{Proceedings of the Third Workshop on Rapid Malcode
  ACM-SIGSAC WORM 2005}, Fairfax, VA,USA, November 2005.

\bibitem{hamhart05}
I.~Hamadeh, J.~Hart, G.~Kesidis, and V.~Pothamsetty, ``A preliminary simulation
  of the effect of scanning worm activity on multicast,'' in \emph{Proceedings
  of the Workshop on Principles of Advanced and Distributed Simulation (PADS)},
  Monterey, CA, USA, June 2005.

\bibitem{will02}
M.~Williamson, ``{Throttling Viruses: Restricting Propagation to Defeat
  Malicious Mobile Code},'' in \emph{Proceedings of the ASAC Security
  Conference}, Las Vegas, NV, USA, 2002.

\bibitem{will03}
J.~Twycross and M.~Williamson, ``{Implementing and testing a virus throttle},''
  in \emph{Proceedings of 12th USENIX Security Symposium}, Washington, DC, USA,
  August 2003.

\bibitem{vecyan03}
G.~Veciana and X.~Yang, ``Fairness, incentives and performance in peer-to-peer
  networks,'' in \emph{Proceedings of the Forty-First Annual Allerton
  Conference on Control, Communications and Computing}, Monticello, IL, USA,
  October 2003.

\bibitem{qiusri04}
D.~Qiu and R.~Srikant, ``{Modeling and performance analysis of BitTorrent-like
  peer-to-peer networks},'' in \emph{Proceedings of the {ACM} {SIGCOMM}},
  Portland, Oregon, USA, August 2004.

\bibitem{masvoj05}
L.~Massoulie and M.~Vojnovic, ``{Coupon Replication Systems},'' in \emph{ACM
  Sigmetrics 2005}, Banff, Alberta, Canada, June 2005.

\bibitem{lilnic04a}
M.~Liljenstam and D.~Nicol, ``{Models of Active Worm Defenses},'' in
  \emph{Proceedings of the IPSI-2004 Studenica Conference}, Studenica, Serbia,
  June 2004.

\bibitem{coh03}
B.~Cohen, ``{Incentives to build robustness in BitTorrent},'' 2003,
  documentation from BitTorrent: http://www.bittorrent.com.

\bibitem{lilyua02}
M.~Liljenstam, Y.~Yuan, B.~J. Premore, and D.~Nicol, ``{A Mixed Abstraction
  Level Simulation Model of Large-Scale Internet Worm Infestations},'' in
  \emph{Proceedings of the Tenth IEEE/ACM Symposium on Modeling, Analysis and
  Simulation of Computer and Telecommunication Systems (MASCOTS)}, Fort Worth,
  TX, USA, October 2002.

\end{thebibliography}
\end{document}